\documentclass{article}

\PassOptionsToPackage{compress}{natbib}

    \usepackage[preprint]{tackling_climate_workshop_style}

\usepackage[utf8]{inputenc} 
\usepackage[T1]{fontenc}    
\usepackage{hyperref}       
\usepackage{url}            
\usepackage{booktabs}       
\usepackage{amsfonts}       
\usepackage{amsmath}
\usepackage{xcolor}         
\usepackage{soul}           
\usepackage{nicefrac}       
\usepackage{microtype}      
\usepackage{float}  
\usepackage{graphicx}
\usepackage{todonotes}
\usepackage{siunitx}
\usepackage{multirow}
\usepackage{tabularx}
\usepackage{subcaption}
\usepackage{placeins}

\title{Technical Report: \\Towards Unified Diffusion Models for Multi-Model Climate Emulation at Scale}

\author{%
  Francesco ~Immorlano\thanks{Denotes equal contributions} \\
  Department of Computer Science\\
  University of California, Irvine\\
  Irvine, CA 92697 \\
  \texttt{fimmorla@uci.edu} \\
  \And
  Elijah ~Tavares\footnotemark[1] \\
  Department of Computer Science\\
  University of California, Irvine\\
  Irvine, CA 92697 \\
  \texttt{eitavare@uci.edu} \\
  \And
  Felix ~Draxler \\
  Department of Computer Science\\
  University of California, Irvine\\
  Irvine, CA 92697 \\
  \texttt{fdraxler@uci.edu} \\
  \And
  Padhraic ~Smyth \\
  Department of Computer Science\\
  University of California, Irvine\\
  Irvine, CA 92697 \\
  \texttt{smyth@uci.edu} \\
  \AND
  Pierre ~Gentine \\
  Department of Earth and Environmental Engineering\\
  Columbia University\\
  New York, NY 10027 \\
  \texttt{pg2398@columbia.edu} \\
  \And
  Stephan ~Mandt \\
  Department of Computer Science\\
  University of California, Irvine\\
  Irvine, CA 92697 \\
  \texttt{mandt@uci.edu} \\
}

\begin{document}

\maketitle

\begin{abstract}

    Large ensembles of climate projections are essential for characterizing uncertainty in future climate and extreme weather events, yet computational constraints of numerical climate models limit ensemble sizes to a small number of realizations per model. We present a unified conditional diffusion model that dramatically reduces this computational barrier by learning shared distributional patterns across multiple Coupled Model Intercomparison Project phase 6 models and emission scenarios. Rather than training separate emulators for each model-scenario combination, our approach captures the common statistical structures underlying nine CMIP6 models, generating daily temperature maps with a global coverage for historical and future periods. This unified framework enables: \textit{(i)} efficient probabilistic sampling for comprehensive uncertainty quantification across models and scenarios; \textit{(ii)} rapid generation of large ensembles that would be computationally intractable with traditional climate models; \textit{(iii)} variance-reduced treatment effect analysis via fixed-seed generation that disentangles forced climate responses from internal variability. Evaluations on held-out models demonstrate reliable generalization to unseen future climates, enabling rapid exploration of different emission pathways.
    
    

\end{abstract}

\section{Introduction}
    The current acceleration of climate change is making extreme weather events more frequent and intense on a global scale, placing new challenges for adaptation and mitigation planning. Robust decision-making requires large ensembles of climate projections to assess risks, guide infrastructure investments, and design resilient systems for the full range of possible future outcomes and their associated uncertainties \cite{ipcc2023climate}. 
    The Coupled Model Intercomparison Project Phase 6 (CMIP6) coordinates a major international effort in which several global climate models provide standardized projections under different Shared Socioeconomic Pathways (SSPs) \cite{eyring2016overview}. However, due to their computational complexity, each simulation requires weeks to months on supercomputers, constraining ensemble sizes to a small number of members per model and severely limiting exploration of initial condition uncertainty and parameter sensitivity \cite{acosta2024copmputational, palomas2025reducing, mahesh2025huge1, mahesh2025huge2}.
    
    Reduced-complexity models, or emulators, were developed to replicate large-scale climate responses more efficiently, but most of them simulate only global-mean variables and lack regional structure  \cite{nicholls2020reduced, nicholls2021reduced, watsonparris2022climatebench}. 
    Recent advances in deep generative models, particularly diffusion models \cite{sohldickstein2015deep, ho2020denoising, nicholdhariwal_2021improved}, offer a promising alternative. These models not only provide orders-of-magnitude computational speedups but also learn complete probability distributions that preserve the statistical properties of climate simulations and enable proper uncertainty quantification through efficient sampling. However, current state-of-the-art approaches emulate individual climate models---requiring separate training for each \cite{bassetti2024diffesm}---and only focus on historical periods \cite{brenowitz2025climate}. 

    We propose a unified conditional diffusion model that simultaneously learns the joint distribution over daily temperature maps from multiple CMIP6 models under different SSP scenarios. Therefore, rather than requiring separate trainings for each model-scenario combination, our model captures the common statistical structures underlying diverse climate models' responses to forcing, enabling flexible generation across nine CMIP6 models and three emission pathways within a single framework.
    
    This study provides three fundamental advances for climate science. First, efficient probabilistic sampling allows comprehensive uncertainty quantification across models and scenarios, capturing complete temperature distributions essential for the characterization of extreme weather events. Second, the approach achieves orders-of-magnitude computational speedup over numerical climate models: generating thousands of daily global temperature fields in minutes rather than months, it makes large-ensemble analyses feasible for impact assessment and adaptation planning. Third, we introduce a \textit{variance-reduced treatment effect estimation} through deterministic sampling: by fixing the random seed across scenarios, we isolate forced climate responses from internal variability improving the statistical efficiency of the treatment effect estimation by 50--300$\times$ compared to standard approaches for equivalent precision.

\section{Multi-Model Climate Emulation}
    Climate simulations exhibit complex dependencies on multiple factors ranging from physical parameterizations to initial conditions and external forcings. For temperature projections, some of these are the specific Earth system model used, the emission scenario, the time of year (therefore, the seasonality), and the temporal evolution of greenhouse gas forcing. Our goal is to train a single diffusion model on temperature simulations from nine CMIP6 models and three SSP scenarios, learning to distinguish their characteristic responses while capturing the underlying probability distribution. This unified approach enables our model to generate temperature distributions for any combination of CMIP6 model, SSP scenario, and time period, providing a computationally efficient alternative to running expensive numerical simulations.
    Critically, because the unified diffusion captures patterns from different CMIP6 models and scenarios, we can rapidly evaluate how different emission pathways affect climate outcomes, enabling policy-relevant exploration at a fraction of the computational cost.

    \subsection{Data}
    To train our unified emulator, we need two complementary datasets: spatial temperature maps simulated by CMIP6 models for historical and future periods and the corresponding greenhouse gas forcing data to condition the generation.
    
    \paragraph{CMIP6 simulation data}
        We use global daily maps simulated over 1850--2100 by nine CMIP6 models (Table \ref{table-s2}) under three emission scenarios: SSPs 2-4.5, 3-7.0, and 5-8.5. Daily resolution is essential for capturing temperature extremes, seasonal cycles, and day-to-day fluctuations that define the full probability distribution our model learns. For each model-scenario combination, we use a single realization (\textit{r1i1p1f1}), regridding all simulations via bilinear interpolation to the CanESM5-CanOE grid specification (64$\times$128 grid points; spatial resolution: $\sim$250 km) to ensure spatial consistency across models.

    \paragraph{CO$_2$ equivalent data}
        To enable temperature map generation under different SSPs, a critical conditioning variable for our model is the global-mean CO$_2$ equivalent (CO$_2$e) concentration, which measures the combined impact of multiple greenhouse gases (CO$_2$, methane, nitrous oxide, etc.) on global warming in equivalent amounts of CO$_2$. Following \cite{meinshausen2020theshared}, we derive these values from radiative forcing estimates simulated by the Minimal CMIP Emulator v1.2 \cite{tsutsui2022minimal} taking into account aerosols and greenhouse gases. This provides one annual CO$_2$e value per year from 1850 to 2100 for each SSPs 2-4.5, 3-7.0, and 5-8.5, matching our CMIP6 simulation periods and scenarios.
        Despite generating daily temperature maps, annual CO$_2$e resolution is appropriate because greenhouse gases are well-mixed in the atmosphere with residence times of years to centuries, making sub-annual and regional variations negligible for climate modeling.
        
    \subsection{Unified conditional diffusion model}
        Our model is trained to learn the conditional distribution $P(T | m,c_s,d,y)$ over daily global temperature maps $T \in \mathbb{R}^{64\times128}$, where the conditioning variables correspond to: 
        \begin{itemize}
            \item $m\in \{0,1\}^{|\mathcal{M}|}$: One-hot encoding of the CMIP6 model to distinguish between different climate models' characteristic responses. 
            \item $c_s, s\in\mathcal{S} = \{\text{SSP2-4.5}, \text{SSP3-7.0}, \text{SSP5-8.5}\}$: Annual global-mean CO$_2$e value for scenario $s$.
            \item $d \in \{0,1\}^{365}$ : One-hot encoding of day of year.
            \item $y \in [1850, 2100]$: Calendar year.
        \end{itemize}
        
        This strategy allows our diffusion approach to distinguish model-specific responses across multiple emission scenarios and generate temperature maps for any specified combination of conditioning factors.
        We condition on both year $y$ and CO$_2$e concentration $c_s$ to enable counterfactual generation across emission scenarios. Although these are correlated within each scenario's trajectory, treating them as independent variables allows the model to generate temperatures for arbitrary year-CO$_2$e combinations. Section \ref{sec:results} validates this capability through variance-reduced treatment effect estimation.


    \subsection{Training Setup}
        \paragraph{Training and test split for generalization assessment}
        To rigorously evaluate our diffusion model's ability to generalize to unseen future climates, we design a hierarchical data split (Table~\ref{table-s1}, Fig.~\ref{fig:fig_data_split}). Of the nine CMIP6 models in our dataset, six are fully included in the training set with both their historical (1850--2014) and SSP simulations (2015--2100). The remaining three--CESM2-WACCM, MPI-ESM1-2-HR, and EC-Earth3-Veg-LR (hereafter CESM, MPI, and EC-Earth)--serve as \textit{test models} for evaluating cross-model generalization. Critically, we include their historical simulations (1850--2014) in the training set, providing the diffusion model with information about each test model's spatial patterns simulated in the historical period. However, we reserve all SSP projections (2015--2100) from test models exclusively for evaluation. Formally, we hold out $\mathcal{M}_{\text{test}} \times \mathcal{S}_{\text{SSP}}$, where $\mathcal{M}_{\text{test}} = \{\text{CESM}, \text{MPI}, \text{EC-Earth}\}$ and $\mathcal{S}_{\text{SSP}} = \{\text{SSP2-4.5}, \text{SSP3-7.0}, \text{SSP5-8.5}\}$.

\section{Evaluation}
    After building the unified diffusion model, we now validate its performance and demonstrate its practical utility for climate analysis. Our model generates independent daily temperature maps without explicitly modeling temporal autocorrelation between consecutive days. This design choice implies that day-to-day aleatoric variability---the inherent unpredictability in weather patterns---is not captured by the model, therefore is treated as noise. This characteristic is reinforced by training across multiple climate models, each exhibiting different internal variability patterns. This implies we cannot assess the diffusion model by comparing individual generated samples against specific ground truth maps, as the chaotic day-to-day fluctuations in CMIP6 simulations are not reproduced. Instead, we must determine whether our model accurately captures the \textit{distributional properties} of CMIP6 simulations that emerge when examining many samples. Our evaluation aims at checking if generated samples reproduce the probability distributions of CMIP6 temperatures, including spatial patterns and seasonal cycles and if our approach generalizes to unseen futures. We focus on MPI as our primary test case, with results for other models reported in the Appendix \ref{subsec:supp_figs_tables}.



\begin{figure}[H]
          \centering
          \includegraphics[width=1\textwidth]{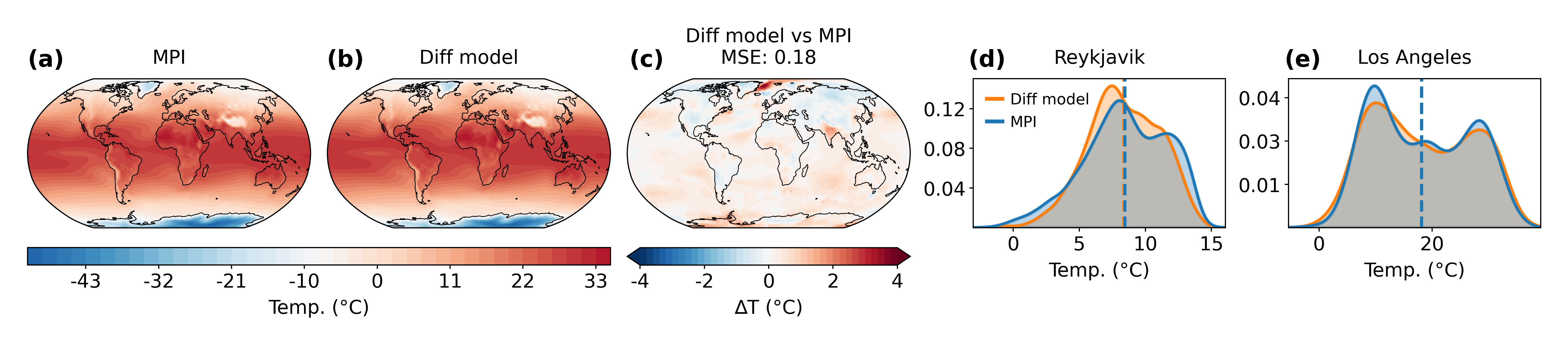}
          \caption{
            (a) Daily temperature maps simulated by MPI and (b) sampled by the diffusion model under SSP3-7.0, averaged over 2070–2080; (c) difference map (diffusion model vs. MPI); (d,e) comparison of daily distributions for the grid points closest to Reykjavik and Los Angeles over 2070--2080.
          }
          \label{fig:MPI_spatial_maps_comparison}
        \end{figure}

    \paragraph{Spatial/temporal patterns and probability distributions}
        We first assess whether generated samples reproduce spatial temperature patterns and probability distributions of held-out CMIP6 simulations. Fig.~\ref{fig:MPI_spatial_maps_comparison} shows results for MPI under SSP3-7.0, averaged over 2070–2080. The generated maps (Fig.~\ref{fig:MPI_spatial_maps_comparison}b) closely reproduce CMIP6 simulations (Fig.~\ref{fig:MPI_spatial_maps_comparison}a), capturing regional temperature patterns. The difference map (Fig.~\ref{fig:MPI_spatial_maps_comparison}c) reveals the largest errors in the Arctic region, likely reflecting the high spatial variability of this area. Beyond spatial patterns, we evaluate the distributional accuracy at specific locations. For Reykjavik and Los Angeles (Fig.~\ref{fig:MPI_spatial_maps_comparison}d,e), generated samples reproduce the complete probability distributions of daily temperatures over 2070–2080, including tail behaviors important for extreme event assessment. This confirms that our model captures the full range of temperature variability. It is worth highlighting that only the historical part (1850--2014) of MPI's simulations was included in the training set (Table~\ref{table-s1}). The model's ability to accurately reproduce the SSP held-out projections contributes to show it has learned meaningful relationships between forcing and climate response from other CMIP6 models, rather than memorizing model-specific trajectories. The same validation for CESM2 and EC-Earth3 is provided in Appendix~\ref{subsec:supp_figs_tables}.

\begin{figure}[h]
        \centering
        \begin{subfigure}[t]{0.45\textwidth}
            \centering
            \includegraphics[width=\textwidth]{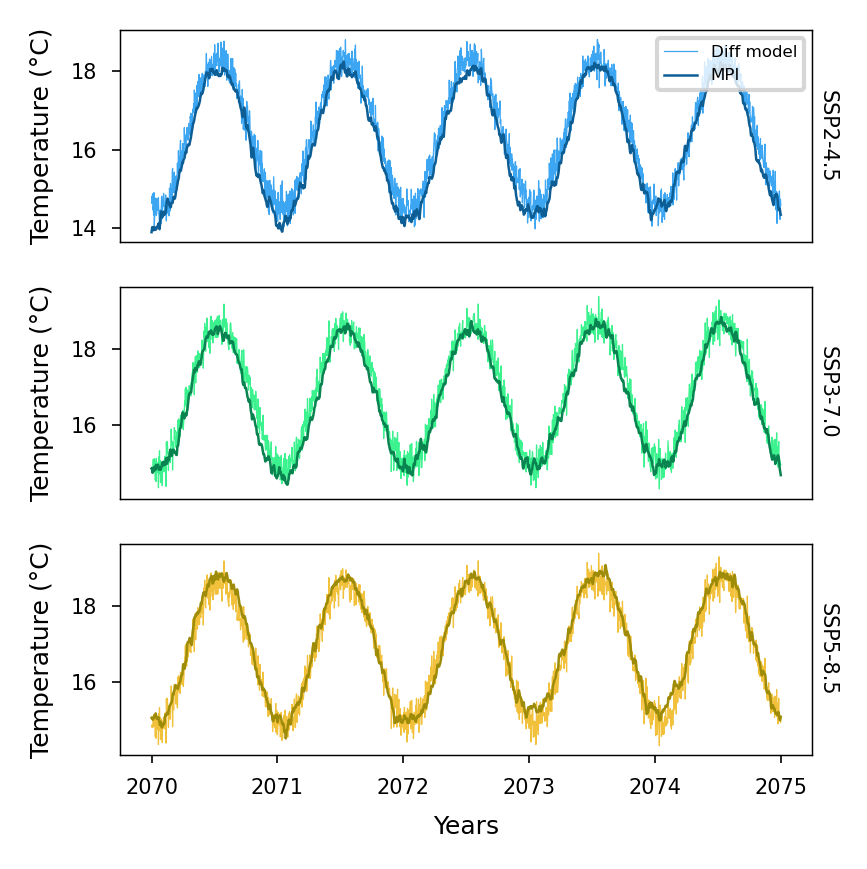}
        \end{subfigure}
        \hfill
        \begin{subfigure}[t]{0.45\textwidth}
            \centering
            \includegraphics[width=\textwidth]{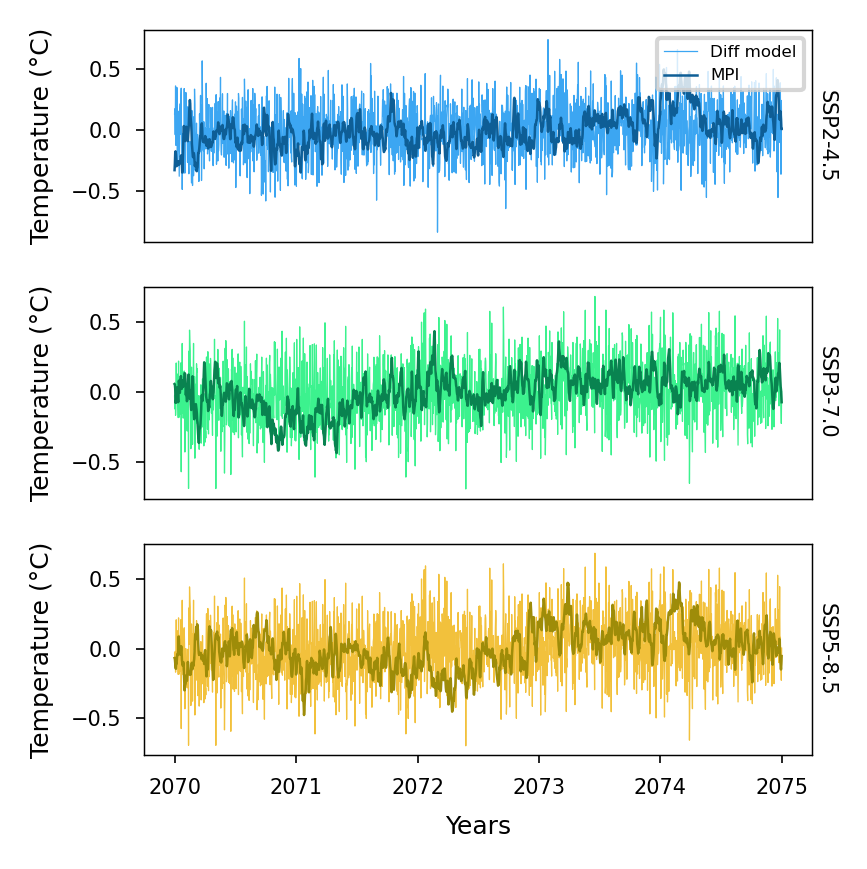}
        \end{subfigure}
        \caption{(a) Global-mean daily temperatures simulated by MPI-ESM1-2-HR and generated by the unified diffusion model over 2070--2075 for the three SSPs used in this work; (b) same as (a), after subtracting the seasonal cycle}
        \label{fig:MPI_global_mean_timeseries}
    \end{figure}

        In addition, Fig.~\ref{fig:MPI_global_mean_timeseries} validates temporal consistency through global-mean daily temperatures over 2070--2075. The generated samples accurately reproduce the seasonal cycle across all SSPs (Fig.~\ref{fig:MPI_global_mean_timeseries}a), matching both amplitude and phase of MPI simulations. After removing the climatological seasonal cycle (Fig.~\ref{fig:MPI_global_mean_timeseries}b)), deseasonalized temperatures reveal realistic day-to-day internal variability around appropriate mean levels. It is worth noting that the diffusion model exhibits larger day-to-day fluctuations than the individual MPI realization: an expected consequence of training on multiple climate models with diverse internal variability patterns.

\begin{figure}[H]
          \centering
          \includegraphics[width=1\textwidth]{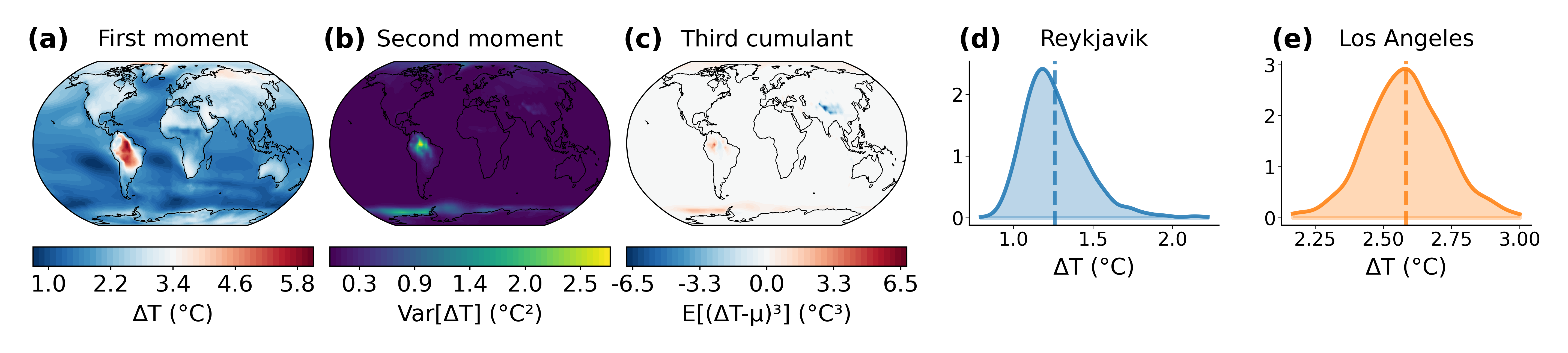}
          \caption{
            Distributional analysis of regional temperature shifts between SSPs 5-8.5 and 2-4.5 for October 17th, 2100, using 1,000 sample pairs generated by the diffusion model emulating MPI. For each sample pair, identical random seeds were used across scenarios to isolate the climate forcing signal. (a-c) Spatial distribution of statistical moments: (a) First moment shows mean temperature change and reveals expected regional warming patterns with Arctic amplification and land-ocean contrasts., (b) Second moment shows variance and quantifies uncertainty of temperature change, with highest values in polar regions and South America indicating less confident predictions; (c) Third cumulant shows asymmetry of temperature change distributions with near-zero skewness globally, confirming symmetric distributions without bias toward extreme outcomes. (d,e) Location-specific probability distributions for Reykjavik and Los Angeles, with Reykjavik's broader distribution reflecting greater uncertainty at high latitudes (dashed lines indicate mean temperatures).
          }
          \label{fig:fig_3}
        \end{figure}

    \paragraph{Statistical Moments}
        Beyond validating spatial and temporal accuracy, we analyze how our diffusion model characterizes uncertainty in climate responses. Figure~\ref{fig:fig_3} examines \textit{treatment effects}, defined as temperature differences between SSPs 5-8.5 and 2-4.5 for October 17th, 2100, using 1,000 sample pairs generated with identical random seeds across SSPs. This allows us to isolate the effect of forcing from stochastic variability (see Section~\ref{paired_seed_treatment_effect_analysis}). The first moment (Fig.~\ref{fig:fig_3}a) reveals expected warming patterns: strongest in Arctic and mid-latitudes regions, weak over oceans, and physically consistent land-ocean contrasts. The second moment (Fig.~\ref{fig:fig_3}b) identifies where the model exhibits greatest uncertainty, concentrated in polar regions and South America. The near-zero third cumulant (Fig.~\ref{fig:fig_3}c) across nearly all locations indicates symmetric temperature change distributions, confirming the model avoids systematic bias toward extremes. Location-specific distributions (Fig.~\ref{fig:fig_3}d,e) show that our unified diffusion exhibits a sharper, more confident projection for Los Angeles compared to Reykjavik's broader distribution.

        \subsection{Comparison to Baseline Climate Emulators}
            In order to establish the value of our diffusion model for climate emulation we need a systematic comparison against simpler, established baselines. Following the ClimateBench framework~\cite{watsonparris2022climatebench}, we compare our unified diffusion model against Gaussian Process (GP) and Linear Pattern Scaling (LPS). This evaluation quantifies whether added model complexity provides meaningful improvements over cheaper alternatives and reveals the specific advantages of learning complete probability distributions rather than point estimates.

            \subsubsection{Metrics}
                We employ two metrics from  ClimateBench \cite{watsonparris2022climatebench} that focus on different aspects of emulators performance.
                \paragraph{Normalized Root Mean Squared Error (NRMSE)}
                Regional climate projections require emulators that capture spatial patterns besides global-mean changes. ClimateBench defines three Normalized Root Mean Squared Error (NRMSE) variants to assess spatial and temporal accuracy \cite{watsonparris2022climatebench}. NRMSE$_s$ measures spatial pattern errors averaged over time:
                    \begin{equation}
                    \text{NRMSE}_s = \dfrac{\sqrt{\langle (|x_{i,j,t}|_t - |y_{i,j,t,n}|_{n,t})^2 \rangle}}{|\langle y_{i,j} \rangle|_{t,n}}
                    \end{equation}
                NRMSE$_g$ measures errors in global-mean temporal evolution:
                    \begin{equation}
                    \text{NRMSE}_g = \dfrac{\sqrt{|(\langle x_{i,j,t} \rangle - \langle |y_{i,j,t,n}|_n \rangle)^2|_t }} {|\langle y_{i,j} \rangle|_{t,n}}
                    \end{equation}
                These combine into NRMSE$_t$, our primary metric:
                    \begin{equation}
                    \text{NRMSE}_t = \text{NRMSE}_s + \alpha \cdot \text{NRMSE}_g
                    \end{equation}
                
                where $x_{i,j,t}$ is the ground truth from CMIP6, $y_{i,j,t,n}$ is the predicted value at grid point $(i,j)$, time $t$, and ensemble member $n$; $\langle \cdot \rangle$ denotes area-weighted global mean; $|\cdot|_{n,t}$ denotes averaging over members and time; and $\alpha=5$ following ClimateBench~\cite{watsonparris2022climatebench}.
                We focus on NRMSE$_t$as it summarizes both spatial and temporal accuracy in a single metric.
                
                \paragraph{Continuous Ranked Probability Score}
                Since generative models can learn probability distributions, we need metrics that evaluate distributional accuracy beyond mean predictions. Continuous Ranked Probability Score (CRPS) \cite{gneiting2005calibrated} compares predicted and observed distributions and is defined as:
                \begin{equation}
                    \text{CRPS} = \int_{-\infty}^{\infty} (\langle F_{i,j,t}(x) \rangle - \langle F_{i,j,t}(y) \rangle)^2 dx
                \end{equation}
                where $F(x)$ and $F(y)$ are the cumulative distribution functions over the predicted and target ensembles, respectively. When observations are single values rather than ensembles (similar to our setup where only one simulation per CMIP6 model is available), the observation $y$ is treated as a degenerate distribution with CDF given by the Heaviside step function $H(x-y)$~\cite{matheson1976scoring, zamo2018estimation}. CRPS then becomes:

                \begin{equation}
                \text{CRPS} = \left\langle \int_{-\infty}^{\infty} [F_{i,j,t}(x) - H(x-y_{i,j,t})]^2 dx \right\rangle_{i,j,t}
                \end{equation}

                where $F_{i,j,t}(x)$ is the empirical CDF from generated samples at location $(i,j)$ and time $t$, and $\langle \cdot \rangle_{i,j,t}$ denotes spatiotemporal averaging.

                Lower CRPS indicates better probabilistic accuracy. Importantly, CRPS reduces to mean absolute error for deterministic forecasts, making it a proper scoring rule that rewards both accuracy and well-calibrated uncertainty \cite{gneiting2005calibrated}.

            \subsubsection{Baseline emulators}
            \paragraph{Gaussian Processes} 
                GP emulators model temperature as a function of forcing with Gaussian-distributed uncertainty, providing closed-form predictive distributions that make them computationally efficient for interpolation within the training range. Training a single GP across multiple models and scenarios is infeasible, as GPs cannot distinguish between different CMIP6 models. Therefore, we train separate GP emulators for each test model on historical (1850--2014) and SSP5-8.5 (2015--2100) data. Training on the full temperature range from historical through the highest emission scenario ensures the GP observes the maximum variability, providing the best possible baseline for comparison.
            
            \paragraph{Linear Pattern Scaling} 
                LPS assumes that regional temperatures scale linearly with global-mean warming. For each climate model and scenario, LPS learns a spatial pattern per degree of global warming during training, then applies this pattern to predict future temperatures. Like GP, we train individual LPS emulators per test model on historical and SSP5-8.5 simulations. As a deterministic method, LPS cannot be evaluated with CRPS.

            \paragraph{Comparison results}
                We evaluate all three approaches on held-out SSP projections (SSP2-4.5 and SSP3-7.0) for our three test models (CESM, MPI, EC-Earth3) across three periods: 2020--2040, 2040--2060, and 2080--2100.It should be noted that our unified diffusion model operates under a more challenging regime: it  emulates all test models simultaneously using only their historical data (1850--2014), having learned multi-model relationships from other CMIP6 models. In contrast, GP and LPS emulators train individually on each test model with full access to its historical and SSP5-8.5 trajectories. 
                Table \ref{table:baselines_comparison} presents quantitative results. For NRMSE$_t$, our model is competitive but does not consistently outperform the baselines which show the lowest errors in most cases. This is expected: GP and LPS perform well at interpolation within their training regime when provided model-specific future data. However, our unified diffusion achieves substantially lower CRPS scores across all model-period combinations, often by factors of 1.3--3$\times$. This advantage is most pronounced in the long-term future (2080--2100) where the test case for MPI shows CRPS values of 1.05 (diffusion) versus 3.32 (GP).

                The divergence between NRMSE$_t$ and CRPS results reveals the value of our approach. NRMSE$_t$ measures only mean prediction accuracy---a metric where well-tuned deterministic emulators naturally excel, especially with model-specific training data. In contrast, CRPS evaluates the entire probability distribution, rewarding both accurate means and well-calibrated uncertainty. This confirms that, while GP can provide basic uncertainty estimates through Gaussian assumptions, it cannot capture the complex distributions of climate variables, particularly for extreme events. Instead, our diffusion---with its CRPS performance---demonstrates it has learned not just average climate responses but also the distributional structure of temperature variability.
        
            \begin{table}[H]
            \centering
            \caption{Our method outperforms baselines on CRPS (NRMSE${_t}$ $\downarrow$, CRPS $\downarrow$). Quantitative evaluation on held-out CMIP6 test models under SSP3-7.0. Our approach achieves consistently lower CRPS scores than both baselines, indicating superior probabilistic accuracy. NRMSE$_t$ measures temporal mean squared error, while CRPS evaluates probabilistic accuracy. Linear Pattern Scaling, being deterministic, cannot produce CRPS scores. Bold indicates best performance for each model-period combination. All metrics computed using 50 samples per day over the specified time periods.}
            \begin{tabularx}{\textwidth}{lXXXXXXXXX}
                \toprule
                & \multicolumn{3}{c}{\textbf{Ours}} & \multicolumn{3}{c}{Gaussian Process} & \multicolumn{3}{c}{Linear Pattern Scaling} \\
                \cmidrule(lr){2-4} \cmidrule(lr){5-7} \cmidrule(lr){8-10}
                  & CESM & MPI & EC-Earth3 & CESM & MPI & EC-Earth3 & CESM & MPI & EC-Earth3 \\
                \midrule
                \multicolumn{10}{c}{\textbf{2020-2040}} \\
                \midrule
                NRMSE$_t$ & 0.0104 & 0.0057 & 0.0073 & 0.0088 & \textbf{0.0050} & \textbf{0.0061} & \textbf{0.0072} & 0.0098 & 0.0095 \\
                CRPS & \textbf{1.13} & \textbf{1.07} & \textbf{1.12} & 1.46 & 1.76 & 1.79 & - & - & - \\
                \midrule
                \multicolumn{10}{c}{\textbf{2040-2060}} \\
                \midrule
                NRMSE$_t$ & 0.0133 & \textbf{0.0063} & 0.0113 & \textbf{0.0079} & 0.0071 & \textbf{0.0061} & 0.0086 & 0.0126 & 0.0099 \\
                CRPS & \textbf{1.17} & \textbf{1.06} & \textbf{1.15} & 1.49 & 2.18 & 2.15 & - & - & - \\
                \midrule
                \multicolumn{10}{c}{\textbf{2080-2100}} \\
                \midrule
                NRMSE$_t$ & 0.0163 & \textbf{0.0065} & 0.0135 & \textbf{0.0188} & 0.0113 & 0.0018 & \textbf{0.0055} & 0.0109 & 0.0118 \\
                CRPS & \textbf{1.3} & \textbf{{1.05}} & \textbf{1.18} & 1.57 & \textbf{3.32} & 3.06 & - & - & - \\
                \bottomrule
            \end{tabularx}
            \label{table:baselines_comparison}
            \end{table}
    \section{Sample-Efficient Treatment Effect Estimation}
        A key benefit of climate emulators is performing uncertainty quantification for treatment effects at a single point in time. Traditional climate models would require rolling out complete trajectories for each realization, simulating daily data from 2015 to 2100 to be comparable to our setup, making large ensembles computationally prohibitive. Our diffusion model samples any point directly, but precise estimation still requires averaging over many samples. We show that by sharing latent noise across conditions, we can estimate the model's learned average treatment effect with $50$--$300\times$ fewer sample pairs.
        \paragraph{Shared Latent Noise Sampling}\label{paired_seed_treatment_effect_analysis} 
            To perform shared latent noise sampling you (1) fix the random seed, (2) generate a latent noise tensor $\mathbf{z} \sim N(0,\mathbf{I})$ in code immediately after fixing the seed, (3) solve the probability flow ODE to obtain a temperature map. Since both the random number generator and ODE solver are deterministic, one seed maps to a single latent gaussian noise tensor\footnote{A random seed initializes the RNG to a deterministic state that produces a stream of random numbers. We draw $\mathbf{z}$ immediately after seeding, before any other stochastic operations, to ensure a one-to-one mapping between seeds and latent tensors.}, which maps to a single temperature map given some fixed conditioning. As a practical benefit, pre-generating latent noise tensors enables batched sampling on a GPU with this technique.
        
            \paragraph{Paired vs Unpaired Seed Sampling} 
                Shared latent noise sampling allows us to pre-generate samples and, when analyzing the data, define treatment effects across any of our conditioning (e.g. switching from 2025 to 2100, SSP2-4.5 to SSP3-7.0, or CESM to MPI) that were sampled using the same latent noise.
        
            This leads to the notion that treatment effects can be computed in two ways using a diffusion model:

            \textit{Paired Sampling}: Generate pairs of temperature maps using identical latent noise, changing a single conditioning variable while holding others fixed, to isolate the effect of that variable.

            \textit{Unpaired Sampling}: Generate pairs of temperature maps using independent latent noise, changing a single conditioning variable, conflating the effect of conditioning with stochastic variability.
        
            \paragraph{Treatment Effects}
                Since each seed determines a unique latent noise tensor, let $s$ denote a seed, with $T_1(s)$ and $T_0(s)$ representing the temperature maps generated with seed $s$ under treatment (e.g., SSP5-8.5) and baseline (e.g., SSP2-4.5) conditioning respectively.

                The \textit{average treatment effect} (ATE) is the expected treatment effect over all possible seeds:
                \begin{align}
                    \text{ATE} = \mathbb{E}_{s}[T_1(s) - T_0(s)]
                \end{align}
        
            \paragraph{Estimating the ATE}
                Given the set of all seeds $\mathbf{s}$, we can estimate the ATE in two ways.
                The \textit{paired estimator} uses the same seed for both conditions:
                \begin{align}
                    \widehat{\text{ATE}}_{\text{paired}} = \frac{1}{N}\sum_{i=1}^{N} 
                    \left[ T_1(\mathbf{s}_i) - T_0(\mathbf{s}_i) \right]
                \end{align}
                The \textit{unpaired estimator} uses different seeds for each condition, where $\mathbf{t}$ is a random permutation of $\mathbf{s}$:
                \begin{align}
                    \widehat{\text{ATE}}_{\text{unpaired}} = \frac{1}{N}\sum_{j=1}^{N} 
                    \left[ T_1(\mathbf{s}_j) - T_0(\mathbf{t}_j) \right]
                \end{align}

\begin{figure}[H]
            \centering
            \includegraphics[width=0.75\linewidth]{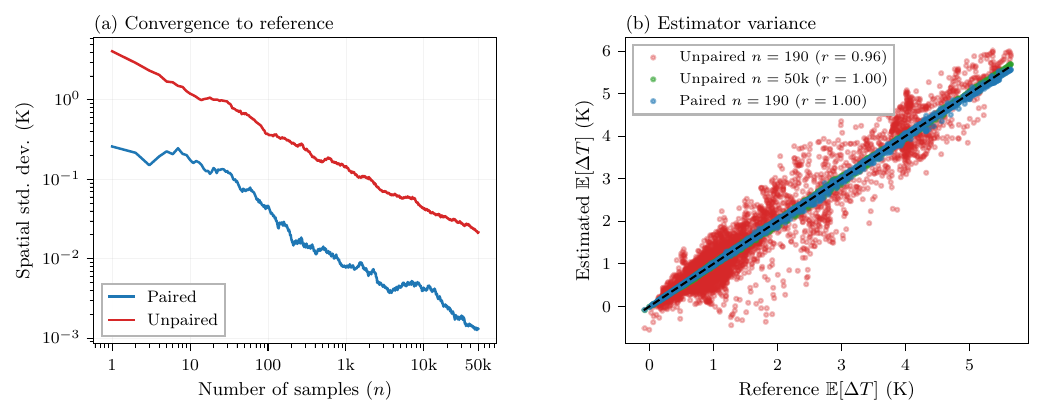}
        \caption{\textbf{Paired vs Unpaired Seeds: Treatment Effect Convergence and Efficiency} 
        (a) Spatial standard deviation of error from the reference ATE map as a function of sample size $n$. 
        The reference is computed from 50,000 held-out paired samples. Treatment effects are computed between 
        SSP5-8.5 and SSP2-4.5 for a single day and year. The paired estimator converges substantially faster, 
        achieving equivalent precision with approximately 50--300$\times$ fewer samples. 
        (b) Scatter plot visualizing estimator accuracy across all grid cells. Each point compares the estimated treatment effect to the reference at one location, with proximity to the diagonal indicating agreement. This shows a full distribution of estimation errors rather than a single summary statistic.
Paired estimation with $n=190$ samples (blue) achieves comparable precision to unpaired estimation with $n=50,000$ samples (green), both achieving $r=1.00$. Unpaired estimation at the same $n=190$ (red) shows greater scatter ($r=0.96$), demonstrating the variance reduction from shared latent noise sampling.}
        \label{fig:convergence-efficiency}
            \end{figure}

                Both estimators converge to approximately the same ATE. However, we empirically observe
                \begin{align*}
                    \operatorname{Var}[T_1(\mathbf{s}_i)-T_0(\mathbf{s}_i)] \ll \operatorname{Var}[T_1(\mathbf{s}_i)-T_0(\mathbf{t}_i)]
                \end{align*}
                as evidenced by the faster convergence of the paired estimator shown in Figure~\ref{fig:convergence-efficiency}. 
                This means the paired estimator achieves comparable precision with far fewer samples. Intuitively, paired seeds isolate the effect of conditioning on the resulting temperature map, whereas unpaired seeds require more samples to average out the random noise.
                Figure~\ref{fig:sample_efficiency_appendix} demonstrates this efficiency gain across sample sizes, with even a single paired sample (n=1) having comparable precision to 241 unpaired samples.

\section{Climate Science Relevant Applications}\label{sec:results}
    By learning to emulate multiple CMIP6 models simultaneously across different SSPs, our approach enables three fundamental advances in climate analysis: probabilistic assessment of future climate states moving beyond deterministic projections, evaluation of probabilistic inter-model differences, rapid cross-scenario exploration for policy insights, and comprehensive uncertainty characterization that maps regional prediction confidence across different climate futures. We demonstrate each capability below.
    \subsection{Learned Distribution Shifts}
\begin{figure}[H]
            \centering
            \includegraphics[width=1\textwidth]{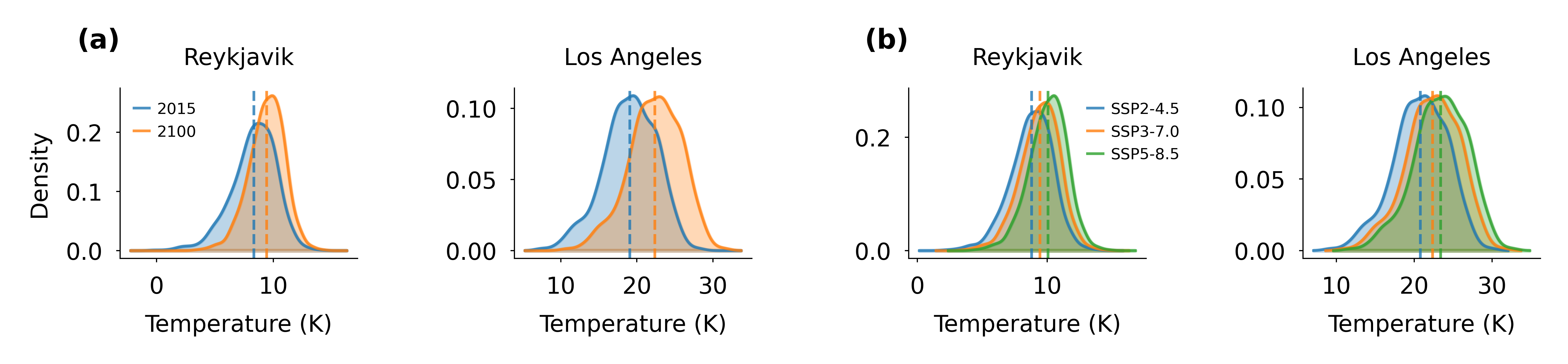}
            \caption{a) Temperature distribution shifts between October 17th, 2015 and 2100 under SSP3-7.0 for the grid points closest to Reykjavik and Los Angeles. b) Temperature distribution shifts across SSP2-4.5, 3-7.0, and 5-8.5 on October 17th, 2100 for the same locations. In both cases, the probability density functions were derived from 1,000 samples generated by the diffusion model emulating MPI, with dashed vertical lines indicating mean temperatures.}
            \label{fig:years_scenarios_distr_shifts}
        \end{figure}

    \paragraph{Probabilistic climate risk assessment beyond mean projections} 
        Fig. \ref{fig:years_scenarios_distr_shifts}a illustrates temperature distribution shifts between 2015 and 2100 for Reykjavik and Los Angeles generated by emulating MPI under SSP3-7.0 (fix \textit{m}, \textit{s}, \textit{d}; vary \textit{c}). These 1,000 samples reveal location-specific climate responses: for example, Los Angeles shows a pronounced rightward shift with increased probability of extreme heat events above 315 K. Such location-specific probabilistic projections are essential for infrastructure planning and risk assessment, providing decision-makers with the full range of possible outcomes.

    \paragraph{Rapid policy scenario comparison at regional scales}
        The unified approach allows instantaneous comparisons of climate outcomes across emission pathways. Fig. \ref{fig:years_scenarios_distr_shifts}b quantifies the distributional impact of different SSPs on the same locations in 2100 (fix $m$, $d$, $c_s$; vary $y$). The progressive rightward shift from SSP2-4.5 to SSP5-8.5 not just translates into differences in mean warming but also into changes in the tail behavior. This rapid scenario comparison is fundamental for policymakers to understand the probabilistic benefits of mitigation strategies at a regional level.

\begin{figure}[H]
        \centering
        \includegraphics[width=1\textwidth]{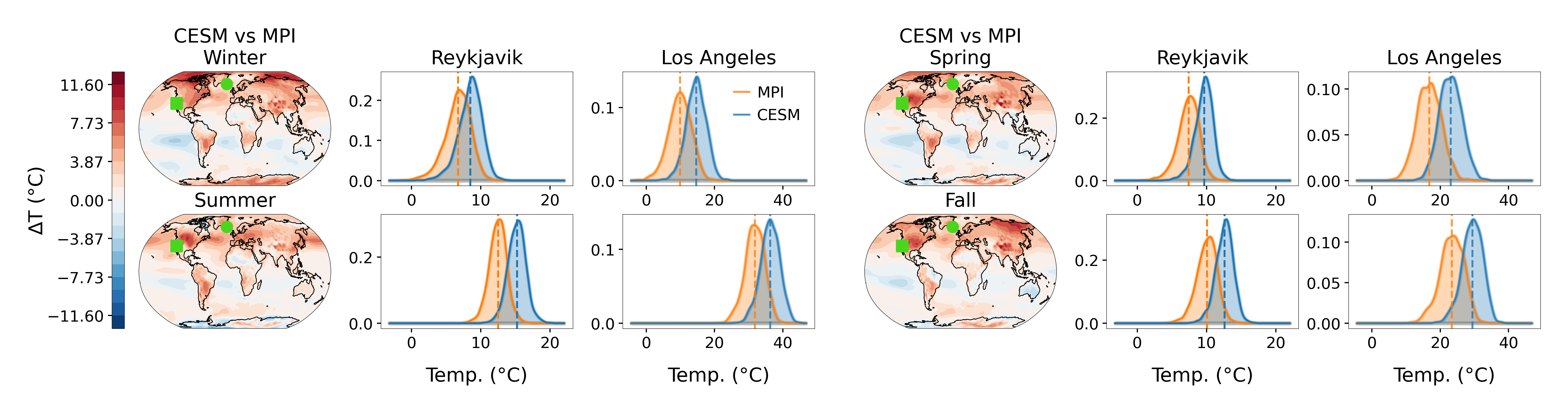}
        \caption{Difference of maps generated emulating CESM and MPI under SSP5-8.5 for: Jan. 31st (Winter), Apr. 20th (Spring), Jul. 19th (Summer), and Oct. 17th (Fall), 2100. Square and circle markers indicate Reykjavik and Los Angeles locations. For each day, temperature distributions for these cities are derived from 1,000 samples, with dashed lines showing mean temperatures.
        }
        \label{fig:seasonal_comparison}
    \end{figure}

    \paragraph{Quantifying distributional differences between climate models}
        Our model captures systematic differences between climate models that persist across scenarios. Fig. \ref{fig:seasonal_comparison} shows how emulations of CESM and MPI differ in their temperature projections across seasons (SSP5-8.5, 2100)---fix $y$, $c_s$; vary $m$, $d$. The spatial patterns show model-dependent biases that vary seasonally with the corresponding distributions that quantify these differences probabilistically for specific locations. Understanding such structural model differences helps interpret the robustness of climate projections and identify regions where model agreement is high and regions where more caution is needed due to model divergence.

\section{Conclusion}
We introduce a unified conditional diffusion model that emulates multiple CMIP6 models and SSPs within a single framework. This provides three main advances for climate science. 

First, rather than requiring separate training for each model-scenario combination, our approach learns statistical patterns underlying diverse climate responses to forcing, enabling flexible generation across CMIP6 models and emission pathways.

Second, benchmarking against GP and LPS demonstrates that while they often exhibit lower NRMSE$_t$ when trained on model-specific future data, our approach achieves substantially lower CRPS scores. This confirms the diffusion model captures not only mean responses but complete temperature distributions, including tail behaviors that simple parametric methods cannot represent. 

Third, we introduce variance-reduced treatment effect estimation through paired-seed sampling. By fixing random seeds across scenarios, we isolate forced climate responses from sampling variability, achieving equivalent statistical precision with 50--300$\times$ fewer samples compared to standard unpaired estimation. This enables efficient exploration of climate outcomes for different SSP scenarios that would be computationally prohibitive with traditional approaches. 

Current limitations include using temperature as the only variable and a spatial resolution of 250 km, which may miss local features critical for impact assessment. Moreover, the diffusion model generates daily maps without explicitly modeling temporal autocorrelation. Despite these constraints, our unified framework enables the generation of thousands of climate projections across models and scenarios at a fraction of the traditional computational cost, providing a practical tool for probabilistic scenario exploration.

\bibliographystyle{unsrt}
\bibliography{references}

\medskip

\small


\appendix

\setcounter{table}{0}
\renewcommand{\thetable}{S\arabic{table}}

\setcounter{figure}{0}
\renewcommand{\thefigure}{S\arabic{figure}}

\section{Appendix / Supplementary Material}\label{sec:appendix}
\subsection{Model architecture}

\begin{figure}[H]
    \centering
    \hspace{-2cm}
    \scalebox{0.45}{%
        \input{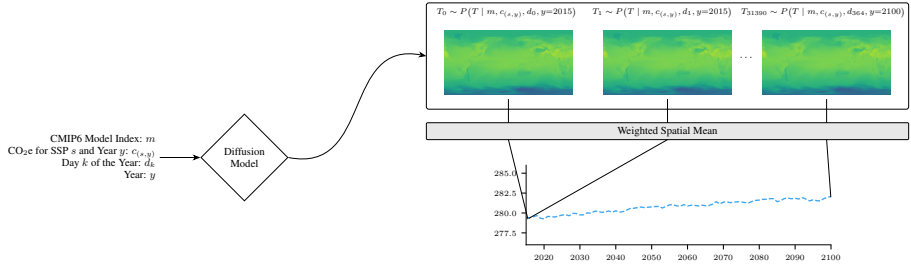}%
    }
    \caption{Architecture of the unified conditional diffusion model. The model takes four conditioning inputs: CMIP6 model index ($m$), CO$_2$ equivalent concentration ($c_{(s,y)}$, determined by SSP scenario $s$ and year $y$), day of year ($d_k$), and year ($y$), to generate daily temperature maps sampling from $P(T \mid m, c_{(s,y)}, d_k, y)$. Each map represents an independent sample generated without temporal correlation between different days. The example shows temperature maps generated for consecutive days from year 2015 to 2100 ($T_0, T_1, \ldots, T_{31389}$; 365 days $\times$ 86 years = 31,390), which when spatially averaged produce the displayed time series. This framework enables generation of large ensembles by sampling multiple realizations for identical conditioning parameters, providing comprehensive uncertainty quantification for any specified model-scenario-time combination.}
    \label{fig:fig_diagram}
\end{figure}

\subsection{Model architecture technical details}
\paragraph{Conditioning}
The conditioning mechanism processes each input according to its type. Categorical variables (CMIP6 model $m$ and day of year $d$) are first one-hot encoded, then transformed through independent sinusoidal positional embeddings to produce $\mathbf{e}_m$ and $\mathbf{e}_d$). Continuous variables (CO$_2$e concentration $c_s$ and year $y$) are processed through separate learned feed-forward networks to compute $\mathbf{e}_c$ and $\mathbf{e}_y$. These four embeddings are concatenated and processed through an additional feed-forward network to create the final conditioning vector:
$$
\mathbf{c}_{\text{comb}} = \text{FFN}([\mathbf{e}_m; \mathbf{e}_c; \mathbf{e}_d; \mathbf{e}_y])
$$
\paragraph{UNet}
Our architecture employs a U-Net backbone with progressive channel expansion (32→64→128 channels using multipliers [1,2,4]) and ResNet blocks at each resolution level. Self-attention is applied at the coarsest resolution. The conditioning vector is projected to match the timestep embedding dimension (128) and combined additively with the timestep embedding throughout the network.

\paragraph{Diffusion Model}

We build on the Elucidating Diffusion Model (EDM) framework \cite{karras2022elucidating}, using a log-uniform noise schedule with $\sigma_{\min}=0.002$, $\sigma_{\max}=200$, and $\rho=7$ following \cite{brenowitz2025climate}. For sampling, we solve the probability flow ordinary differential equation using Heun's second-order method with 50 steps, which provides deterministic generation given an initial latent noise vector. This deterministic sampling is critical for our treatment effect analysis (see Section \ref{paired_seed_treatment_effect_analysis}), as it allows generating paired samples that differ only in conditioning variables while holding the random seed fixed.

\subsection{Data}
\paragraph{Data Pre-processing}
We standardize temperature maps using \textit{z-score normalization} based on training set statistics mean $\mu = \SI{279.87}{\kelvin}$  and standard deviation  $\sigma = \SI{21.14}{\kelvin}$. After standardization, we set $\sigma_{\text{data}}=1$. Note that the original temperature distribution exhibits substantial left skewness (z-scores ranging from approximately -5 to +2), which is preserved after standardization.
        

\paragraph{Loss}
We modified the EDM loss to weight each grid cell by the cosine of its latitude, ensuring that the loss contributions are proportional to actual geographic area. This correction is standard practice when working with regular latitude-longitude grids to avoid over-optimization for high-latitude regions. Indeed, in such grids, polar regions occupy the same number of pixels as equatorial ones, despite representing much smaller areas on Earth's surface.

\begin{figure}[H]
    \centering
    \includegraphics[width=0.75\linewidth]{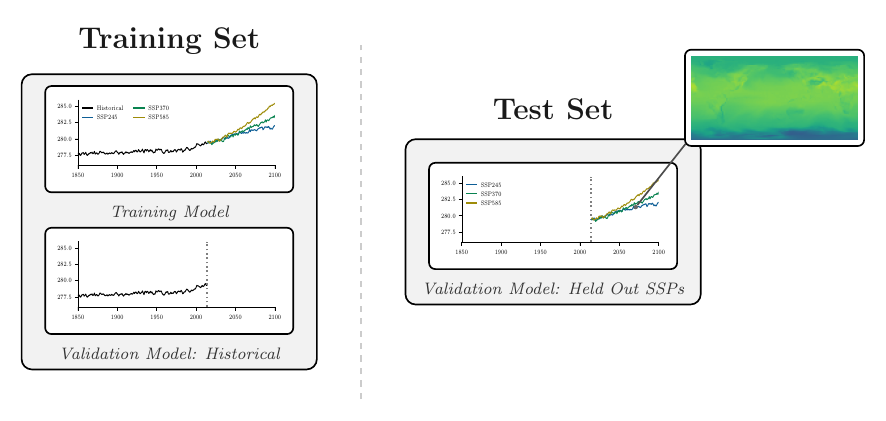}
    \caption{Data split strategy for diffusion model training illustrated with global-mean temperature time series. Six CMIP6 models (\textit{Training Models}) contribute both historical (1850--2014) and SSP simulations (2015--2100) to the training set ("Hist+SSPs" models in Table \ref{table-s2}), shown with complete temperature trajectories under all scenarios. Three models serve as validation cases where only historical data through 2014 (indicated by dotted line) is included in training, while their SSP projections (2015--2100) are withheld for testing the model's ability to generalize to unseen future scenarios ("Hist" models in Table \ref{table-s2}. The sample temperature map (top right) illustrates the spatial data that underlies each time series point.}
    \label{fig:fig_data_split}
\end{figure}

\begin{table}[H]
  \caption{Data split for training and test sets. \textit{Hist} refers to the historical simulation (1850--2014) of the corresponding model. \textit{SSPs} refers to the SSP2-4.5, 3-7.0, and 5-8.5 simulations (2015--2100) of the corresponding model.}
  \label{table-s1}
  \centering
  \begin{tabular}{lll}
    \toprule
    Model        & Train set    &     Test set   \\
    \midrule
    AWI-CM1-1-MR & \textit{Hist} + \textit{SSPs} &  \\
    CanESM5     & \textit{Hist} + \textit{SSPs} &   \\
    CESM2-WACCM     & \textit{Hist} & \textit{SSPs}  \\
    EC-Earth3-Veg-LR     & \textit{Hist}  &  \textit{SSPs} \\
    IPSL-CM6A-LR     & \textit{Hist} + \textit{SSPs} &  \\
    MIROC6     & \textit{Hist} + \textit{SSPs}  &     \\
    MPI-ESM1-2-HR     & \textit{Hist}   &  \textit{SSPs} \\
    MRI-ESM2-0     & \textit{Hist} + \textit{SSPs} &   \\
    NorESM2-MM     &  \textit{Hist} + \textit{SSPs} &  \\
    \bottomrule
  \end{tabular}
\end{table}

\subsection{Supplementary figures and tables}\label{subsec:supp_figs_tables}

\begin{table}[H]
  \caption{List of CMIP6 models used in this work.}
  \label{table-s2}
  \centering
  \begin{tabular}{ll}
    \toprule
    Model         \\
    \midrule
    AWI-CM1-1-MR \cite{semmler2020simulations}  \\
    CanESM5    \cite{swart2019thecanadian} \\
    CESM2-WACCM     \cite{gettelman2019thewhole}       \\
    EC-Earth3-Veg-LR     \cite{doscher2022theec}       \\
    IPSL-CM6A-LR     \cite{boucher2020presentation}       \\
    MIROC6     \cite{kataoka2020seasonal}       \\
    MPI-ESM1-2-HR     \cite{muller2018ahigher}       \\
    MRI-ESM2-0     \cite{yukimoto2019themeteorological}       \\
    NorESM2-MM      \cite{seland2020overview}      \\
    \bottomrule
  \end{tabular}
\end{table}

\begin{figure}[H]
    \centering
    \includegraphics{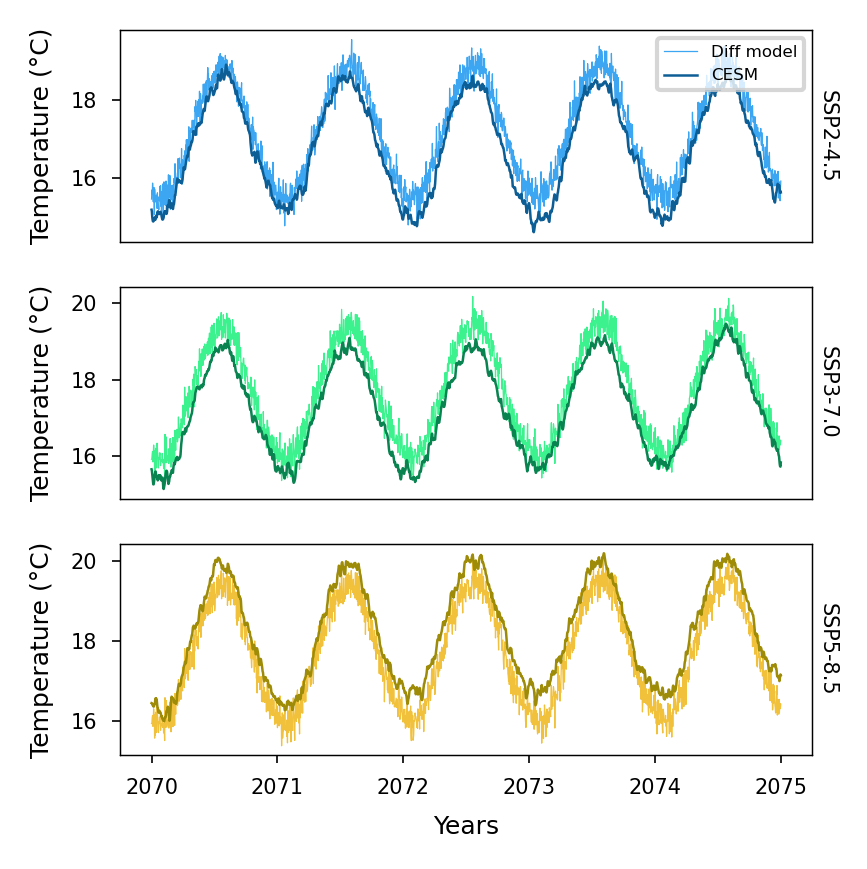}
    \caption{Global-mean daily temperatures simulated by CESM2 and generated by the trained diffusion model over 2070--2075 for the three SSPs used in this work.}
    \label{fig:fig_s1}
\end{figure}

\begin{figure}[H]
    \centering
    \includegraphics{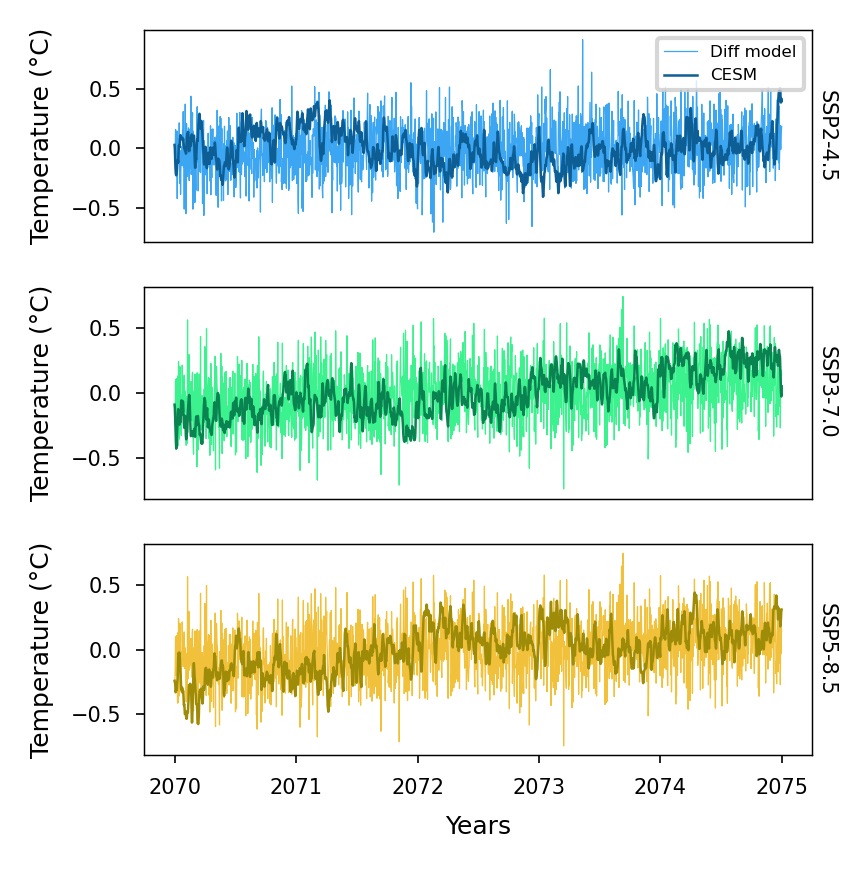}
    \caption{Same as Fig. \ref{fig:fig_s1} after subtracting the seasonal cycle.}
    \label{fig:fig_s2}
\end{figure}

\begin{figure}[H]
    \centering
    \includegraphics{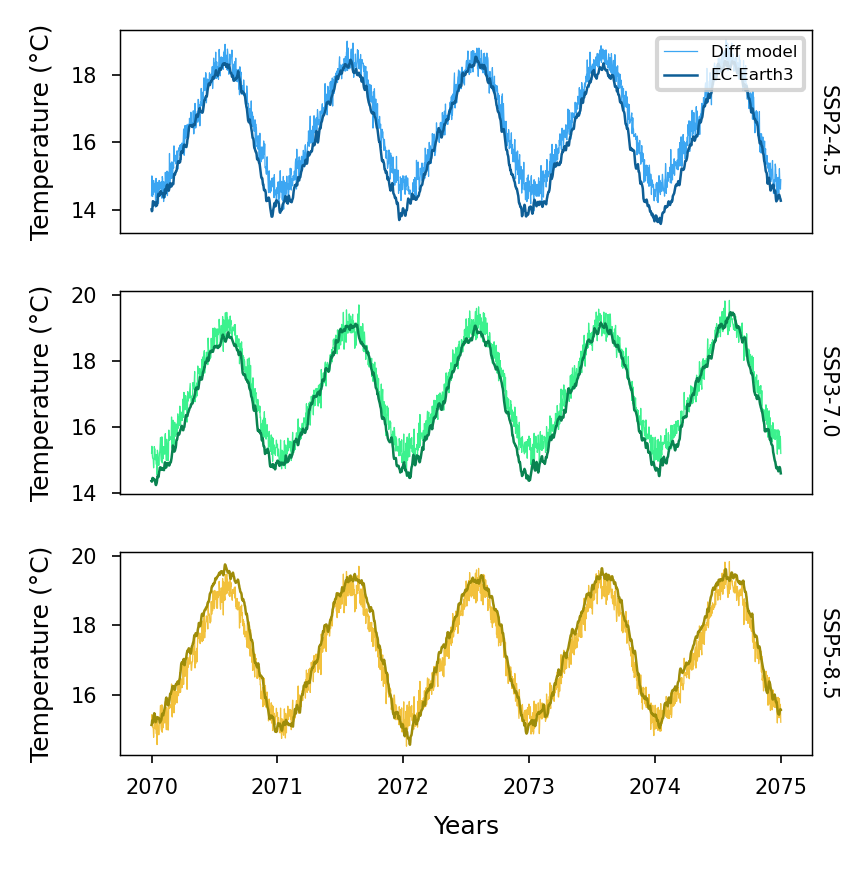}
    \caption{Global-mean daily temperatures simulated by EC-Earth3 and generated by the trained diffusion model over 2070--2075 for the three SSPs used in this work.}
    \label{fig:fig_s3}
\end{figure}

\begin{figure}[H]
    \centering
    \includegraphics{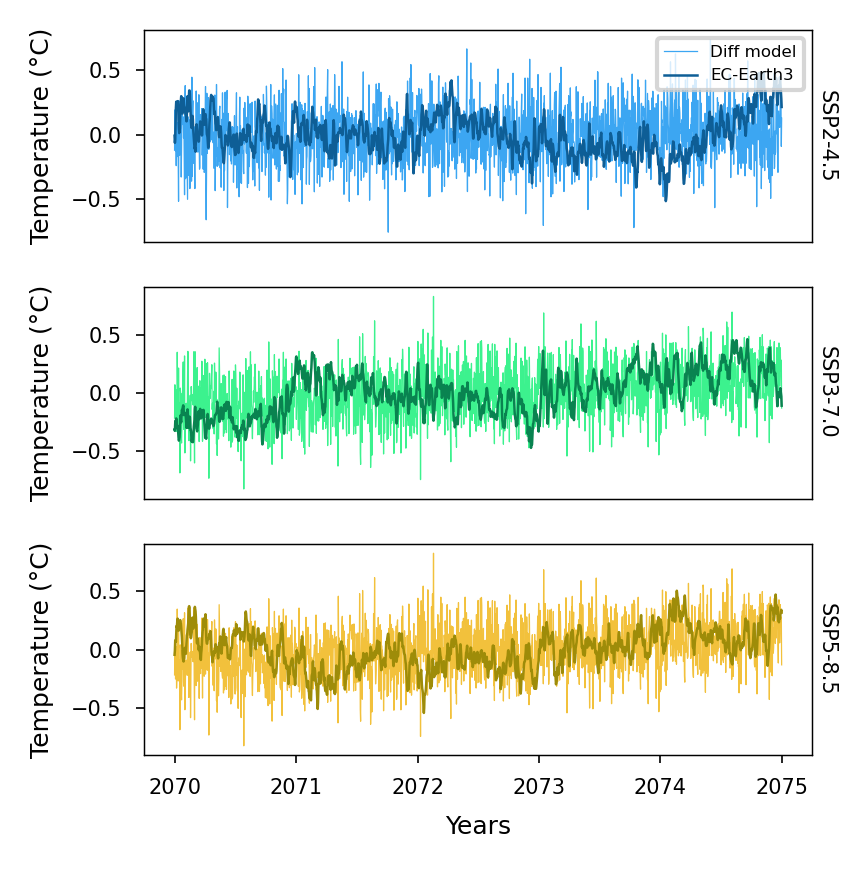}
    
    \caption{Same as Fig. \ref{fig:fig_s3} after subtracting the seasonal cycle.}
    \label{fig:fig_s4}
\end{figure}

\begin{figure}[H]
    \centering
    \includegraphics[width=1\textwidth]{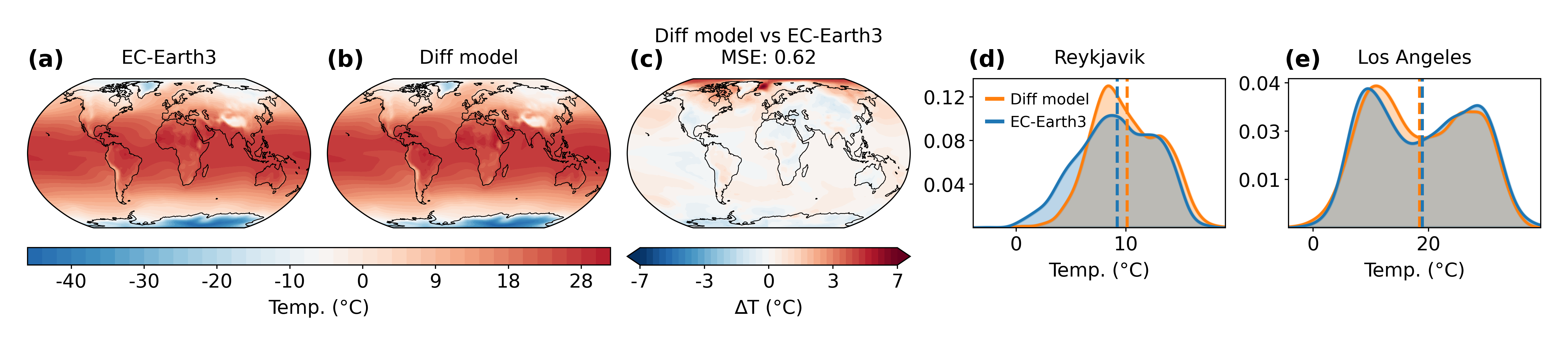}
    \caption{(a) Daily temperature maps simulated by CESM and (b) sampled by the diffusion model under SSP3-7.0, averaged over 2070–2080; (c) difference map (diffusion model vs. CESM); (d,e) comparison of daily probability distributions for the grid points closest to Reykjavik and Los Angeles over 2070–2080.}
    \label{fig:fig_s7}
\end{figure}

\begin{figure}[H]
    \centering
    \includegraphics[width=1\textwidth]{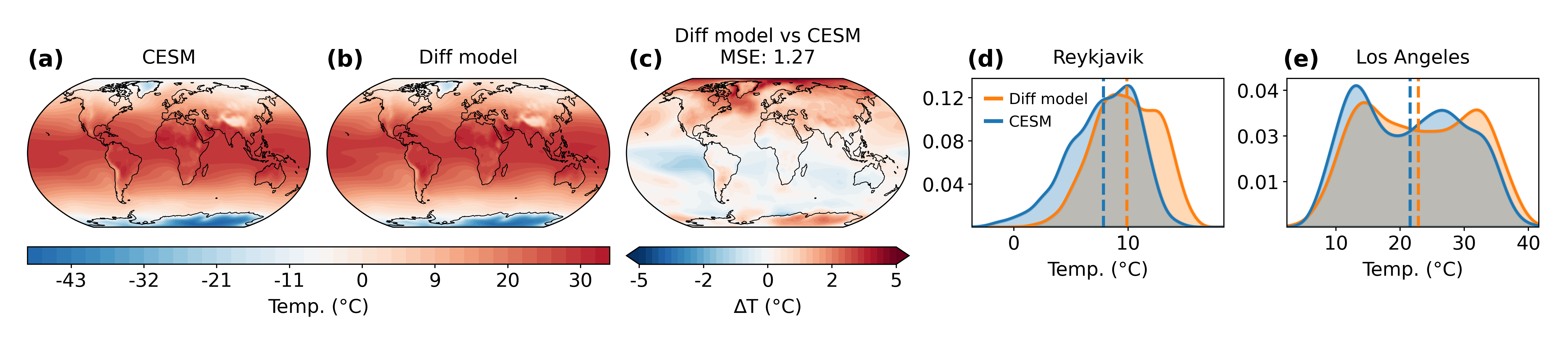}
    \caption{(a) Daily temperature maps simulated by CESM and (b) sampled by the diffusion model under SSP3-7.0, averaged over 2070–2080; (c) difference map (diffusion model vs. CESM); (d,e) comparison of daily probability distributions for the grid points closest to Reykjavik and Los Angeles over 2070–2080.}
    \label{fig:fig_s8}
\end{figure}

\begin{figure}[H]
        \centering
        \includegraphics[width=\linewidth]{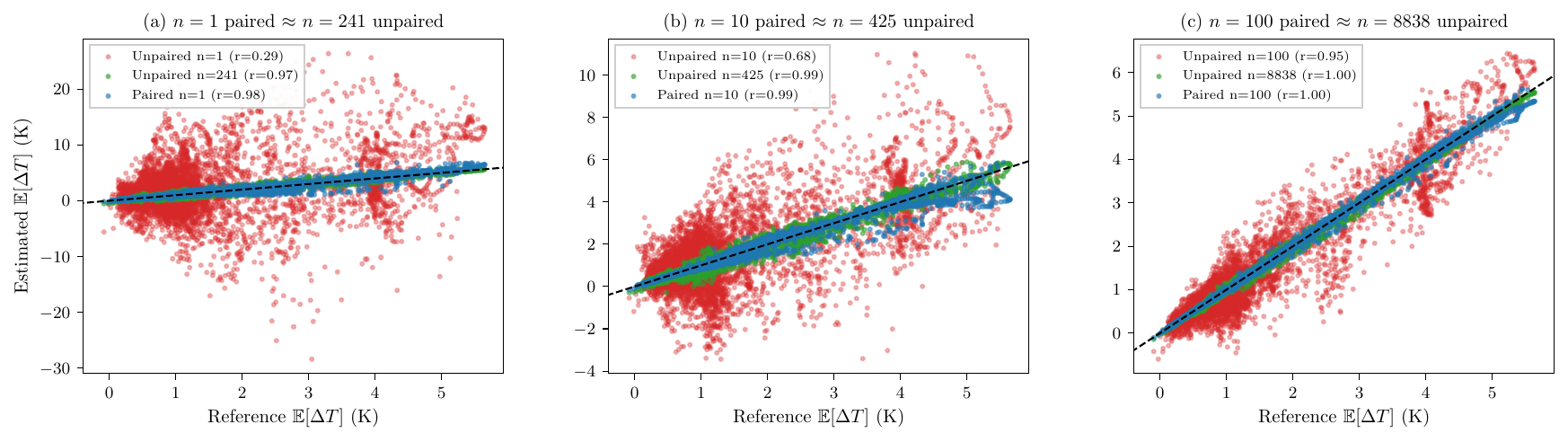}
        \caption{\textbf{Paired Estimates Achieve Comparable Precision with Far Fewer Samples} \newline
        Scatter plots comparing estimated $\mathbb{E}[\Delta T]$ at each grid cell against the reference (50,000 paired samples) for three sample sizes. In each panel: red shows unpaired estimation at $n$ samples, blue shows paired estimation at $n$ samples, and green shows unpaired estimation at the matched sample size required to achieve comparable precision. 
(a) $n=1$: a single paired sample ($r=0.98$) achieves comparable precision to 241 unpaired samples ($r=0.97$). 
(b) $n=10$: paired ($r=0.99$) matches unpaired at $n=425$ (42$\times$ efficiency). 
(c) $n=100$: paired ($r=1.00$) matches unpaired at $n=8838$ (88$\times$ efficiency). 
The tight overlap between blue and green in all panels confirms both estimators converge to the same value, while the scatter in red demonstrates the variance reduction achieved by paired sampling.}
        \label{fig:sample_efficiency_appendix}
    \end{figure}



\end{document}